\documentclass[aps,showpacs,twocolumn]{revtex4}
\usepackage{epsfig}
\usepackage{times}
\usepackage[T1]{fontenc}

\newcommand{\beq}{\begin{equation}}
\newcommand{\eeq}{\end{equation}}
\newcommand{\bea}{\begin{eqnarray}}
\newcommand{\eea}{\end{eqnarray}}

\begin{document}

\title{Phonon Squeezing in a Superconducting Molecular Transistor }
\author{A. Zazunov $^{a)b)c)}$, D. Feinberg$^{b)}$ and T. Martin$^{c)}$}
\affiliation{$^{a)}$ Laboratoire de Physique et Mod\'elisation des Milieux Condens\'es, 
Universit\'e Joseph Fourier, BP 166, 38042 Grenoble, France}
\affiliation{$^{b)}$ Laboratoire d'Etudes des Propri\'et\'es Electroniques
des Solides, CNRS, BP 166, 38042 Grenoble, France}
\affiliation{$^{c)}$ Centre de Physique Th\'eorique, CNRS and Universit\'e de la M\'editerran\'ee, 
Case 907 Luminy, 13288 Marseille Cedex 9, France}

\date{\today}
\begin{abstract}
Josephson transport through a single molecule or carbon nanotube is 
considered in the presence of a local vibrational mode coupled to the electronic charge. 
The ground-state solution is obtained exactly in the limit of a large superconducting gap,
and is extended to the general case by variational analysis. 
The Josephson current induces squeezing of the phonon mode, which is controlled by the superconducting phase difference and by the junction asymmetry. 
Optical probes of non-classical phonon states are briefly discussed.

\end{abstract}


\maketitle

Nanoelectromechanical systems (NEMS) offer a way to reach the quantum regime of mechanical oscillators. A challenging goal consists in creating non-classical vibrational states, similar to non-classical states of light\cite{non-classical,yuen}, 
with reduced quantum fluctuations in one of the mode quadratures. 
The generation of such states has been suggested in bulk materials \cite{polaron,nori}.
Recently, squeezing scenarios have been suggested for resonators driven by nonlinear couplings
\cite{squeezing} and for a Cooper-pair box coupled to a cantilever \cite{armour}.    
Here we consider a superconducting molecular transistor, where the molecule carries both electronic and vibrational degrees of freedom. We show that squeezing 
occurs when a Josephson current flows through the molecule,
by exploiting the coherent regime of polaron dynamics \cite{holstein}.
Compared to the frequency range of micro-cantilevers (less than 1 GHz), here the frequency range goes from 1 GHz to 10 THz depending 
on the system (single wall carbon nanotube (SWCNT) or single molecule). 

The excitation of molecular vibrations (phonons) by an electronic current 
has been observed with normal metallic leads in several molecules including fullerenes and carbon 
nanotubes \cite{metal-molecule}. The latter are in fact NEMS, where charge fluctuations couple to bending 
\cite{bending}, stretching \cite{stretching} or radial breathing (RBM) modes \cite{RBM}. 
Incoherent polaron-like charge fluctuations due to transport trigger a non-equilibrium distribution 
of the phonon mode \cite{theory}. 
In the present paper, we address the superconducting regime. Recently, the proximity effect 
and the Josephson effect \cite{Josephson-nt} have been observed in gated carbon nanotubes. 
Yet, in the tunneling regime and for strong enough electron-phonon ($e$--$ph$) interaction, the Josephson current should be suppressed due to the Franck-Condon dressing factor \cite{novotny}. 
In contrast, we show here that for 
transparent lead-molecule contacts, the Josephson effect survives a strong $e$--$ph$ interaction and triggers coherent phonon fluctuations. As a striking consequence, the conjugate momentum of the molecular 
distortion displays reduced zero-point fluctuations. 
The superconducting phase difference
and the junction asymmetry allow to control non-classical 
squeezed phonon states, including nearly (gaussian) minimum-uncertainty states. 

The model Hamiltonian reads $H = H_M + H_L + H_R + H_T$, 
where $H_M$, $H_{L,R}$ and $H_T$ respectively describe the uncoupled molecule, 
two superconducting leads and the molecule-lead coupling. 
Explicitly (we put $\hbar=k_B=1$),  
\begin{equation}
H_M = \left[ \epsilon_0 - \lambda \left( b + b^\dagger \right) \right] 
\sum_{\sigma = \uparrow, \downarrow} (n_\sigma-{1 \over 2})  + 
U n_\uparrow n_\downarrow + \Omega \, b^\dagger b \,, 
\end{equation}
where $n_\sigma = d_\sigma^\dagger d_\sigma$ with $d_\sigma$, $d_\sigma^\dagger$ 
being the fermion operators for spin $\sigma = \uparrow, \downarrow$ 
on the molecular level $\epsilon_0 = \epsilon_0 (V_g)$,
the position of which can be tuned by the gate, $U$ is the repulsive Coulomb interaction, 
and $b, b^\dagger$ are the boson operators for the local mode of frequency $\Omega$.
The leads are described by standard BCS Hamiltonians ($j=L,R$ is the lead index),
\begin{equation}
H_j = \sum_k \Psi^\dagger_{jk} \left( \xi_k \sigma_z + \Delta \sigma_x \right) \Psi_{jk}~,~~~
\Psi_{jk} = \left(
\begin{array}{c}
\psi_{jk, \uparrow} \\
\psi^\dagger_{j(-k), \downarrow}
\end{array}
\right)~,
\end{equation}
with the energy dispersion $\xi_k$; the Pauli matrices $\sigma_{x,z}$ act in Nambu space. 
The molecule-lead coupling is given by
\begin{equation}
H_T = \sum_{jk} \Psi^\dagger_{jk}~{\cal T}_j ~d + {\rm H.c.}~,~~~
d = \left(
\begin{array}{c}
d_\uparrow \\ d^\dagger_\downarrow
\end{array} \right)~,
\end{equation}
\noindent 
where ${\cal T}_{j=L/R} = t_j \sigma_z ~e^{\pm i \sigma_z \phi / 4}$,
$t_j$ is the $j$th lead-molecule tunneling amplitude, 
and $\phi$ is the superconducting phase difference across the molecule.
Averaging over the leads yields 
a partition function for the molecular site,  
$Z[\phi] = {\rm Tr} \, \left\{ e^{- \beta H_M} W(\beta, 0) \right\}$, with
\begin{equation}
W(\beta, 0) = T_\tau \, \exp \left\{ - \int_0^\beta d \tau d \tau' \,
d^\dagger(\tau) ~\Sigma (\tau - \tau') ~ d(\tau') \right\} \,,
\label{W}
\end{equation}
\noindent
where $T_\tau$ is the imaginary-time ordering operator, 
$\beta$ is the inverse temperature,
$\Sigma (\tau - \tau') =
\sum_j {\cal T}_j^\dagger \, g(\tau - \tau') \, {\cal T}_j$,  and
$g(\tau) = - \sum_k \left( \partial_\tau + \xi_k \sigma_z + 
\Delta \sigma_x \right)^{-1} \delta(\tau)$ is the Green function of the uncoupled leads.
Assuming a constant normal-state density of states in the leads, $\nu$, one obtains:
\begin{equation}
\Sigma(\tau - \tau') = \Gamma \left[
\partial_\tau + \Delta \left( \cos {\phi \over 2}~ \sigma_x + \gamma \sin {\phi \over 2}~ \sigma_y\right)
\right] Q(\tau - \tau') ~,
\label{Sigma_pauli}
\end{equation}
with $Q(\tau) = \beta^{-1} 
\sum_{\,\omega_n} \, e^{- i \omega_n \tau}/\sqrt{\omega_n^2 + \Delta^2}$,
where $\omega_n = \pi (2n + 1)/\beta$ is a fermionic Matsubara frequency
(integer $n$),
$\Gamma = \Gamma_L + \Gamma_R$, 
$\gamma = \left( \Gamma_L - \Gamma_R \right)/\Gamma$, and $\Gamma_j =  \pi \nu t_j^2$.

First we focus on the case where the relevant molecule dynamics is 
restricted to the low-frequency domain on the scale of the superconducting gap 
(in particular, $\Gamma \ll \Delta$). 
In this limit, the $\Sigma$ term in Eq. (\ref{W}) becomes local in time with
$Q(\tau - \tau') \rightarrow \Delta^{-1} \delta(\tau - \tau')$ at low frequencies,
$\omega_n \ll \Delta$. This yields an effective Hamiltonian 
for the molecule as
\begin{equation}
\nonumber
H_{eff} = H_M + 
\Gamma \, d^\dagger \left( \cos {\phi \over 2} \, \sigma_x + 
\gamma \sin {\phi \over 2} \, \sigma_y \right) d \,.
\end{equation}
The Hilbert space of $H_{eff}$ splits into two subspaces, ${\cal H}_1$ and ${\cal H}_2$,
spanned by electron states $\left\{ \, | \uparrow \rangle , | \downarrow \rangle \, \right\}$
and $\left\{ | 0_d \rangle , | \uparrow \downarrow \rangle \right\}$, respectively. 
In the large $\Delta$ limit, these subspaces are decoupled. 
We hereafter assume the repulsion $U$ to be relatively small,
$U/2 < \Gamma$, consistently with good contacts.
Therefore, the ground state of the system lies in the ${\cal H}_2$ subspace \cite{LY}.
Introducing the Pauli-matrix operators
$\tau_z = | 0_d \, \rangle \langle \, 0_d | - 
| \uparrow \downarrow \rangle \, \langle \uparrow \downarrow |$ and
$\tau_{+} = \left( \tau_{-} \right)^\dagger =
| 0_d \rangle \, \langle \uparrow \downarrow |$,
and performing the rotation  
$H_{eff} = e^{i \tau_z \chi/2 } \tilde{H} \, e^{- i \tau_z \chi/2 }$ with
$\chi = \arctan \left(\gamma \tan{\phi \over 2}\right)$
lead to an effective spin-boson Hamiltonian:
\begin{equation}
\tilde{H} = \Omega \, b^\dagger b -  
\left[ \, \epsilon  - \lambda \left( b + b^\dagger \right)  \right] \tau_z  + 
\tilde{\Gamma}(\phi) \, \tau_x  ~,
\label{H2prime}
\end{equation}
where $\tilde{\Gamma}(\phi) = \rho(\phi) \Gamma$,  
$\rho(\phi) = \sqrt{ \cos^2 (\phi/2) + \gamma^2 \sin^2 (\phi / 2)}$, and
$\epsilon = \epsilon_0 + U /2$. 
Any eigenstate of $\tilde H$, and particularly
the ground state $| \tilde \Psi_0 \rangle$, can be written as
\begin{eqnarray}
| \tilde \Psi_0 \rangle = e^{- i \tau_z \chi/2 } | \Psi_0 \rangle = 
\sum_{n = 0}^\infty \left( A^{(0)}_n | 0_d  \rangle +  
A^{(2)}_n | \uparrow \downarrow \rangle  \right) | n  \rangle \,,
\label{gs}
\end{eqnarray}
where $| n  \rangle$ is the $n$-phonon Fock state. 
The state (\ref{gs}) exhibits entangled 
charge and vibrational states. 
Note that for $\epsilon=0$, 
the amplitudes in Eq. (\ref{gs}) fulfill $A_n^{(0)} = (-1)^{n+1} A_n^{(2)}$,
owing to the parity symmetry  
$\left[ \tilde{H} \, , \, \tau_x (-1)^{b^\dagger b} \right] = 0$.

The Josephson current flowing through the molecule can be obtained
as a functional derivative of the generating functional 
$Z_\xi \equiv Z[\phi + \xi(\tau)]$ 
with respect to the source variable $\xi(\tau)$,
$J = -2 \left. \delta \ln Z_\xi / \delta \xi(0) \right|_{\xi=0}$.
After averaging over the leads and neglecting retardation of the kernel
$\Sigma$ in Eq. (\ref{W}), the Josephson current reduces to 
$J = Z^{-1} {\rm Tr} \left\{ e^{- \beta H_M} W(\beta, 0) \hat{J} \right\}$,
where $\hat{J} = \Gamma d^\dagger 
\left( -\sin(\phi/2) \, \sigma_x + \gamma \cos(\phi/2) \, \sigma_y \right) d$
is an effective current operator \cite{AQ} in terms of $d$.
In the rotated basis of Eq. (\ref{H2prime}), the current operator takes the form
\begin{equation}
\tilde J = \left( \Gamma / 2 \rho(\phi) \right)
\left[ \left( 1 - \gamma^2 \right) \sin \phi ~ \tau_x + 2 \gamma \, \tau_y \right] \,.
\label{J}
\end{equation} 
Note that in the presence of phonons, 
the current operator does not commute with the Hamiltonian 
(\ref{H2prime}). This results in quantum fluctuations of the current 
even in the case of $\epsilon = \gamma = 0$. 

In the ground state, the relevant experimental quantity is the current expectation value, 
$J(\phi)=\langle \Psi_0 | \hat{J} | \Psi_0 \rangle$, which probes 
the overlap of phonon states in Eq. (\ref{gs}). 
To make further progress, we perform a unitary transformation 
$U_\alpha = e^{-i \alpha P \tau_z}$ 
on $\tilde{H}$ of Eq. (\ref{H2prime}), where $P=i (b^\dagger-b)$ and
$\alpha = \lambda/\Omega$:
\begin{equation}
{\tilde H'} = \Omega b^\dagger b \,- \,
\epsilon\tau_z  + \tilde{\Gamma}
\left[ \cos(\alpha P) \,  \tau_{x} + \sin(\alpha P) \, \tau_{y} \right] \,,
\label{Hpol}
\end{equation}
To lowest order in $\Gamma/\Omega$, for $\epsilon=0$, one obtains
the ground state  
$| \Psi_0 \rangle= e^{i\chi/2}| \alpha \rangle 
| \, 0_d  \rangle - e^{-i\chi/2}| -\alpha  \rangle 
| \uparrow \downarrow \rangle$, 
where $| \pm \alpha  \rangle = U_{\pm \alpha} |0_{ph}\rangle$ 
are coherent states of the phonon mode. 
In the limit of large $\alpha$, 
the Josephson current 
$J(\phi) = \Gamma e^{-\alpha^2} |\sin (\phi / 2)|$ 
becomes strongly suppressed \cite{novotny}. 
As we show below, in a molecular transistor with a relatively strong coupling to the leads 
a sizeable Josephson current may flow and generate squeezing
without the necessity of an external drive. 
This is in contrast with Ref. \cite{armour} where an external 
oscillator ($\Omega < 1 $GHz) is coupled to a Cooper-pair box, 
for which $\Omega \ll \Gamma$ and $\alpha < 1$ holds.

In order to probe the generic squeezing properties of the molecular phonon, 
we now compute the  mean square fluctuations
of both the displacement coordinate $X=b+b^\dagger$ and the conjugate momentum $P$
in the ground state of the system assuming zero temperature.
The ground state (\ref{gs}) is found numerically 
by truncating the phonon Fock space to a maximum number of phonons,
$n_{max}\gg\lambda/\Omega$. 
In the range of parameters considered, 
$n_{max}=30$ is sufficient to ensure 
the convergence of the calculations. When discussing 
the numerical results, we set $\Omega = 1$.

Coherent charge fluctuations enhance the fluctuations of $X$,  
$\delta X=\left( \langle \Psi_0 | X^2 | \Psi_0 \rangle - 
\langle \Psi_0 | X | \Psi_0 \rangle^2 \right)^{1/2}$,
beyond the quantum zero-point magnitude. Concomitantly, the charge fluctuations 
{\it reduce} the fluctuations of the momentum, $\delta P=\langle \Psi_0 | P^2 | \Psi_0 \rangle^{1/2}$. 
Note that $\langle \Psi_0 | P | \Psi_0\rangle=0$ always, 
while $\langle \Psi_0 | X | \Psi_0 \rangle=0$ only for $\epsilon=0$. 

The effective polaronic interaction constant 
$\lambda^2/\tilde{\Gamma}(\phi)$ of the spin-boson 
Hamiltonian (\ref{H2prime}) depends on the superconducting phase difference 
and becomes infinite at $\phi=\pi$ for a symmetric junction, $\gamma=0$. 
Fig. 1a shows the $\phi$-dependent variation of the momentum fluctuation $\delta P$
as well as the uncertainty $\delta X \delta P$.
(The ground-state value of $\delta X \delta P$ for an harmonic oscillator is $1$.) 
Squeezing ($\delta P<1$) occurs for a wide range of parameters, 
and its intensity depends on $\phi$, 
in accordance with the effective polaronic interaction. 

For small $\Gamma$ (strong $e$--$ph$ interaction), the resulting coherent states 
$| \pm \alpha  \rangle$ are not squeezed, 
i.e., squeezing marks a deviation from the displaced oscillator (coherent) states. 
The generation of gaussian squeezing can be understood 
from the Hamiltonian (\ref{Hpol}) by noticing that $P^2$ is 
not modified by the transformation $U_\alpha$. 
Expanding the third term in Eq. (\ref{Hpol}) for small $\lambda$ gives 
${1\over 2}\alpha^2 \Gamma \left(b^2+(b^\dagger)^2\right)\tau_x $, which precisely generates two-phonon squeezed states \cite{yuen}. 
The variation of $\delta P$ between $\phi=0$ and $\pi$, 
where $\tilde{\Gamma}(\phi)$ is maximum/minimum, 
shows that squeezing is a crossover phenomenon and is maximal for intermediate coupling 
parameters of the polaron problem \cite{polaron}. In general,
squeezing does not involve minimum-uncertainty states.
Yet, $\delta X \delta P=1$ can be made very close to unity
for intermediate $\lambda$ and $\Gamma$ (see Fig. 1b). 
\begin{figure}[h]
\scalebox{0.38}{\includegraphics{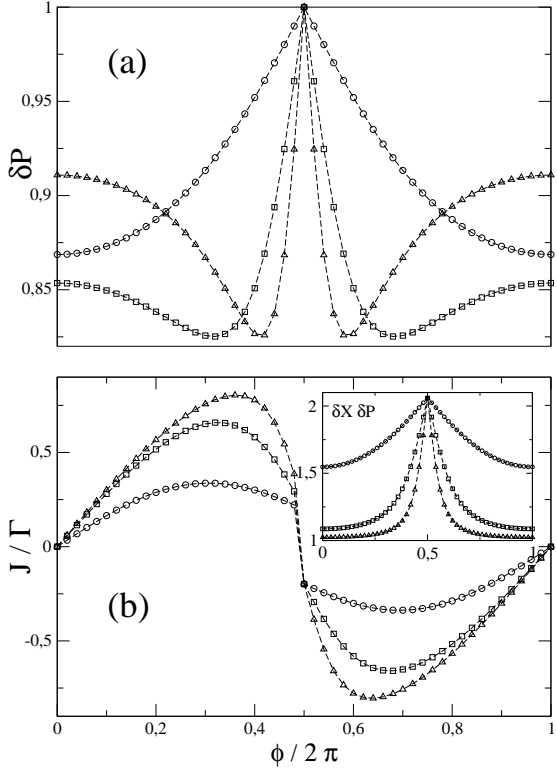}} 
\caption{(a) Squeezing of the momentum and (b) Josephson current as functions of 
the phase difference $\phi$. 
We take $\Gamma = 0.5$ (circles), $2$ (squares), and $4$ (triangles) for $\lambda=0.9$, 
$\epsilon = \gamma=0$.
The inset in (b) shows the Heisenberg uncertainty as a function of $\phi$.}
\label{fig1}
\end{figure}
\begin{figure}[h]
\scalebox{0.38}{\includegraphics{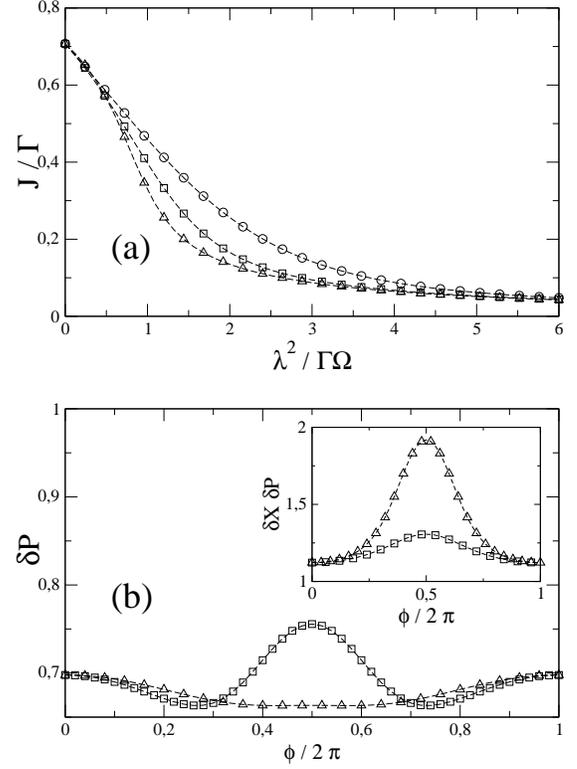}}
\caption{ (a) Josephson current as a function of the electron-phonon interaction, 
at $\phi= \pi/2$: $\Gamma=0.5$ (circles), $1$ (squares) and $2$ (triangles) 
for $\epsilon=\gamma=0$;
(b) Same as Fig. 1a, in the asymmetric case: 
$\gamma=0.5$ (squares) and $0.75$ (triangles) for $\lambda=1.5$, $\Gamma=4$, $\epsilon = 0$.
}
\label{fig2}
\end{figure}

Squeezed states can only be produced with a sizeable Josephson current. 
Actually, optimal squeezing (in the sense of minimum-uncertainty) is generated 
in the parameter range where $J$ is moderately affected by $e$--$ph$ interaction.
To illustrate this, the Josephson current is plotted as a function of the bare interaction 
$\lambda^2/\Gamma\Omega$ (Fig. 2a): as expected, $J$ decreases but moderately if $\lambda$ is not too large. 
The inflexion region corresponds to the polaron crossover where optimum squeezing is achieved. 
Interestingly, the asymmetry $\gamma \neq 0$ leads to strong (up to $40$ percent) and 
nearly harmonic squeezing, having a weak dependence with the phase difference (Fig. 2b).

To check that the squeezing property is generic and 
not only restricted to the large $\Delta$ limit, we have treated
the case of finite $\Delta$ by using a variational ansatz approach \cite{polaron}. 
A modified polaronic transformation 
$U_p(\eta)=e^{\eta \alpha (b^\dagger - b) \tau_z}$ is performed on the original Hamiltonian, 
Eqs. (1)--(3), followed by a harmonic squeezing transformation 
$U_{sq}(\beta)=e^{\beta (b^{\dagger2}-b^2)}$, where $\eta$ and $\beta$ are variational parameters.
Averaging over the zero-phonon state leads to renormalized tunneling amplitudes 
${\cal T}_j'={\cal T}_j\,\exp\left(-\frac{1}{2} \alpha^2 \eta^2 e^{-4 \beta} \right)$
and the effective interaction $U'=U - 2 \eta(2-\eta) (\lambda^2/\Omega)$. 
Squeezing is measured by $\delta P = e^{-2\beta}$.
For large $\Gamma$, using the Hartree approximation, 
one minimizes the ground state energy with respect to $\eta$ and $\beta$. 
This yields a good qualitative agreement with the exact large $\Delta$ calculation. 
For finite $\Delta$, squeezing is still present provided that $\lambda > \Omega$. 
For instance, with the parameters of Fig. 2b 
($\lambda=1.5, \Gamma=4, \gamma=0.75$) and $\phi=0$, 
one finds $\delta P=0.79$ and  0.76 for $\Delta=100$ and 4, respectively. 
In the opposite limit of $\Delta \ll \Gamma$, 
assuming $\Delta=0.1$, $\lambda=1.2$, $\Gamma=2$, $\gamma=0$, 
one finds (at $\phi=\pi/2$) $\delta P=0.77$ with $\delta X \delta P = 1.1$, 
indicating that even with rather high frequency modes such as RBM in nanotubes 
($\Omega \approx 10 meV$), squeezing could be obtained 
(assuming Nb contacts, $\Delta \approx 1 meV$).


Nonclassical phonon states (other than squeezed states) can be generated by projecting the ground 
state on the current eigenstates, which are in principle accessible experimentally. 
The ground state in the current state basis (defined from Eq. (\ref{J}) as 
$\tilde J |\pm \rangle = \pm J |\pm \rangle$) reads:
$| \Psi_0 \rangle = | \Phi_+  \rangle  | + \rangle + | \Phi_- \rangle  | - \rangle$,
with $| \Phi_{\pm}  \rangle \,= \,e^{\mp i\chi' /2} 
\sum_{n = 0}^\infty \left(e^{i\chi/2}A^{(0)}_n  \pm 
e^{-i\chi/2} A^{(2)}_n \right)| n  \rangle$ 
and $\chi' = \arctan \left(\gamma \cot{\phi \over 2}\right)$.
In the small $\Gamma$ regime, one obtains, for $\epsilon=0$, the ``phase cat'' states,
\begin{equation} 
| \Phi_{\pm} \rangle = 
e^{\mp i\chi' /2} \left(e^{i\chi /2}| \alpha \rangle \pm 
e^{-i\chi /2}|-\alpha \rangle \right)/ \sqrt{2} \,,
\end{equation} 
which are superpositions of opposite coherent states of the phonon mode. 
More generally, states like $| \Phi_{\pm}  \rangle$ are similar to those generated in Quantum Optics 
\cite{non-classical}. 
They carry nontrivial phases, obtained here by tuning the junction asymmetry $\gamma$.
In particular, for $\gamma=0$, they are built from phonon Fock states with even/odd occupation numbers only, 
being linked, respectively, to the $| \pm \rangle$ current states. 
At moderate $\lambda$ it is possible to achieve such nontrivial states.
They have a substantial overlap between phonon states and a sizable Josephson current.  
 
We have also checked that squeezing is robust against environmental effects induced 
by coupling the local vibrational mode to an external phonon bath,
by using the reduced density matrix formalism. 
Provided that the quality factor $Q = \Omega/\eta$, 
with $\eta$ the dissipative coupling constant, 
is large enough ($Q\sim10^2-10^4$ has been suggested \cite{bending,stretching,RBM}), 
the squeezing property of the ground state is protected due to the gap 
($\sim {\rm min} ( \Omega, \Delta) $) in the excitation spectrum of the system.

A direct probe of squeezed (non-classical) phonon states is needed. 
For low-$\Omega$ bending modes in SWCNT, 
capacitive detection through a single-electron transistor can be envisioned.
Optical detection techniques may be used, such as Resonant Raman Scattering (RRS), 
which has been achieved in carbon nanotubes \cite{raman} for RBM modes. 
In RRS, the incident photon excites an electronic transition within the molecule, 
and a photon is re-emitted with excitation of the phonon modes. 
We denote $d'$ the state corresponding to a low-lying molecular 
orbital, which is assumed to be decoupled from the molecular vibrations. 
The molecule Hamiltonian then reads:
\begin{equation}
H_M'=H_M+ \epsilon' d'^\dagger d'+\left( \zeta e^{-i\omega t} d^\dagger d' + H.c.\right) \,.
\end{equation}
Using the golden rule to calculate the Raman transition rate,
we assume that the initial state is the projection of the ground state (\ref{gs}) 
on the zero-electron subspace, 
$|i\rangle = |\Phi^{(0)}\rangle |0\rangle_d {|\uparrow \downarrow \rangle}_{d'}$,
with the phonon part  
$| \Phi^{(0)} \rangle=\sum_{n = 0}^\infty  A^{(0)}_n | n  \rangle$, 
while the final state with {\it one} electron in the orbital $d$ 
(together with a hole in $d'$),
involves an arbitrary number of phonons without a net 
displacement, $|f\rangle = |n \rangle |\sigma \rangle_d |-\sigma \rangle_{d'}$. 
Resonant Raman emission lines (Stokes) will appear at energies 
$\epsilon_0-\epsilon' + n\Omega- E_0(\phi)$ 
($E_0(\phi)$ is the ground state energy),
with probabilities given by $\langle n | \Phi^{(0)} \rangle^2=| A_n^{(0)} |^2$; this  yields 
a spectroscopy of the squeezed state $| \Phi^{(0)} \rangle$. 
While absorption from a filled molecular orbital ($d'$) probes $| \Phi^{(0)} \rangle$, similarly absorption towards an empty orbital will probe 
$| \Phi^{(2)} \rangle=\sum_{n = 0}^\infty  A^{(2)}_n | n  \rangle$ 
by taking out an electron from the state $| \uparrow \downarrow \rangle_d$. 
Concerning the "cat states", experimental evidence could be gained in principle by 
far-infrared optical absorption involving a transition from 
the subgap (Andreev) bound states to quasiparticle states in the leads. 

In summary, we have studied theoretically the generation of phonon squeezing 
by coherent charge fluctuations in a superconducting molecular junction. 
Squeezing occurs for a wide range of parameters 
and is maximal in the polaron crossover regime.
A nearly minimum-uncertainty state with about 40 per cent squeezing can be obtained
provided that the local vibrational mode is weakly coupled to the environment. 
For these purposes, suspended nanotubes present the advantage that their various vibrational modes are well characterized and the corresponding quality factors are large. 
Finally, we have discussed the possibility of using optical techniques 
to detect the spectral properties of non-classical phonon states. 

The authors acknowledge support from the A.C. Nanosciences Program NR0114 of Ministry of Research.

\end{document}